\documentclass[intlimits,twoside,a4paper]{article}
\usepackage{amsmath,amssymb}
\usepackage{graphicx}
\usepackage{wrapfig}
\usepackage[T2A]{fontenc}
\usepackage[cp1251]{inputenc}
\usepackage[eqsecnum]{cmpj}

\issue{2011}{14}{2}{23701}

\doinumber{10.5488/CMP.14.23701}

\title[A comparative study for structural and electronic properties]%
{A comparative study for structural and electronic properties of
single-crystal ScN}
\author[R. Mohammad, \c{S}. Kat{\i}rc{\i}o\u{g}lu]{R. Mohammad\refaddr{label1,label2},
        \c{S}. Kat{\i}rc{\i}o\u{g}lu\refaddr{label2}}
\addresses{
\addr{label1} Palestine Technical University, Applied Science
College, WestBank, Palestine \addr{label2} Middle East Technical
University, Physics Department, 06530~Ankara, Turkey}
\authorcopyright{R. Mohammad, \c{S}. Kat{\i}rc{\i}o\u{g}lu, 2011}
\date{Received November 4, 2010, in final form February 3, 2011}

\begin{document}

\maketitle

\begin{abstract}
A comparative study by FP-LAPW calculations based on DFT within
LDA, PBE-GGA, EV$_{\rm ex}$-PW$_{\rm co}$-GGA, and EV$_{\rm
ex}$-GGA-LDA$_{\rm co}$ schemes is introduced for the structural
and electronic properties of ScN in RS, ZB, WZ, and CsCl phases.
According to all approximations used in this work, the RS phase is
the stable ground state structure and makes a transition to CsCl
phase at high transition pressure. While PBE-GGA and EV$_{\rm
ex}$-PW$_{\rm co}$-GGA's have provided better structural features
such as equilibrium lattice constant and bulk modulus, only
EV$_{\rm ex}$-PW$_{\rm co}$-GGA and EV$_{\rm ex}$-GGA-LDA$_{\rm
co}$'s have given the non zero, positive indirect energy gap for
RS-ScN, comparable with the experimental ones. The indirect band
gap of ScN in RS phase is enlarged to the corresponding measured
value by EV$_{\rm ex}$-PW$_{\rm co}$-GGA+U$^{\rm SIC}$
calculations in which the Coulomb self and exchange-correlation
interactions of the localized d-orbitals of Sc have been corrected
by the potential parameter of U. The EV$_{\rm ex}$-PW$_{\rm
co}$-GGA calculations have also provided good results for the
structural and electronic features of ScN in ZB, WZ, and CsCl
phases comparable with the theoretical data available in the
literature. EV$_{\rm ex}$-PW$_{\rm co}$-GGA and EV$_{\rm
ex}$-PW$_{\rm co}$-GGA+U$^{\rm SIC}$ schemes are considered to be
the best ones among the others when the structural and electronic
features of ScN are aimed to be calculated by the same
exchange-correlation energy approximations.
\keywords ScN, FP-LAPW, DFT, structural properties, electronic
properties
\pacs 71.15.Mb, 71.15.Nc, 71.20.Nr, 71.18.+y

\end{abstract}

\section{Introduction}
ScN (Scandium nitride) has many potential applications due to its
high mechanical strength~\cite{1}, good thermal stability with a
melting temperature above $2000^{\rm o}$~C~\cite{2} and high
hardness of 21~GPa with respect to load deformation~\cite{1}.
After the epitaxial growth of smooth and singly oriented ScN
films~\cite{1,2,3,4,5,6,7,8}, ScN was also considered to be a
potential semiconductor in electronic device applications. X-Ray
diffraction analysis indicated that, unlike the other nitride
semiconductors, all ScN films crystallize into a single rock-salt
(RS) phase (B1) with a lattice constant of $\sim$4.5~\AA~\cite{1,
3,4,5,6,7,8}. The RS structure was also determined to be the most
stable structure for ScN by the comparative theoretical
works~\cite{9,10,11}. The lattice constant of ScN in RS phase has
been calculated to be in the range of
4.42--4.651~\AA~\cite{9,11,12,13,14,15} by the first principles
calculations within local density approximation (LDA)~\cite{16,
17} and generalized gradient approximation (GGA)~\cite{18,19} of
exchange and correlation energies. In these
works~\cite{9,10,11,12,13,14}, the bulk modulus of ScN has been
reported in the range of 196--235~GPa with respect to the
experimental value of 182$\mp$40~\cite{1}. In the literature the
most stable RS phase with the cohesive energy of --13.69~\cite{9}
and --13.428~eV~\cite{10} was found to be followed by the wurtzite
(WZ or B4), zinc-blende (ZB or B3), and CsCl (B2) type meta-stable
structures of ScN with the ordered cohesive energies of
--13.35~\cite{9}, --13.03~\cite{9}, and --11.58~eV~\cite{9}
(--11.34~eV~\cite{10}), respectively. These cohesive energies
correspond to the lattice constants of 3.49 ($c/a=1.6$,
$u=0.38$)~\cite{9}, 4.88~\cite{9} and 2.81~\AA~\cite{9}
(2.79~\AA~\cite{10}) for WZ, ZB, and CsCl phases of ScN,
respectively. In a recent work~\cite{11}, the lattice constants
have been calculated to be 3.45, 4.939, and 2.926~\AA \, for ScN
in WZ, ZB and CsCl phases. The bulk modulus of WZ phase of ScN has
been reported to be 156~\cite{9} and 132.91~GPa~\cite{11} by GGA
and LDA of exchange and correlation energies, respectively. In the
same works, the bulk modulus of CsCl phase was calculated to be
170~\cite{9} and 159.759~GPa~\cite{11}. In another work~\cite{10},
the CsCl meta-stable phase of ScN has the bulk modulus of
178.60~GPa by GGA calculations.

The first principles electronic band structure calculations within
LDA~\cite{9,11,12,14} and GGA~\cite{9,10,13} schemes indicated
that RS-ScN was a semi-metal with an almost zero indirect band
gap. However, ScN in RS phase was a semiconductor with the
indirect gap of 2.4~\cite{8}, 1.3$\mp$0.3~\cite{20}, and
0.9$\mp$0.1~eV~\cite{7} measured by optical transmission,
reflection and absorption experiments, respectively. The recent
ab-initio calculations~\cite{14,20,21,15,12} using
pseudopotential~(PP) method within Hedin's~\cite{22} Green's
functions quasiparticle corrections on LDA (LDA-G$_{\rm o}$W$_{\rm
o}$), exact exchange of LDA~\cite{23} [OEPx(cLDA)], and Hedin's
Green's functions quasiparticle corrections on exact exchange of
LDA [OEPx(cLDA)-G$_{\rm o}$W$_{\rm o}$], Linear Muffin Tin Orbital
(LMTO) method within Bechstedt's~\cite{24} Green's functions
quasiparticle corrections on LDA (LDA-GW) and
full-potential-linearized augmented plane waves (FP-LAPW) method
within GGA~\cite{19} and screened exchange of LDA
(sx-LDA)~\cite{25} have all supported the semiconductor nature of
the stable ScN by giving the indirect gap in the range of
0.54--1.70~eV at X symmetry point (E$_{\rm g}^{\rm {\Gamma-X}}$).
In~\cite{12,14,20,21}, the optical direct gap of RS structure at X
point (E$_{\rm g}^{\rm {X-X}}$) was calculated in the range of
1.98--2.90~eV with respect to the experimental direct absorptions
in the range of 1.8--2.4~eV~\cite{3,4,7,8,20}.

The recent DFT calculations within GGA scheme~\cite{9} have
indicated that the WZ and ZB meta-stable structures of ScN were
non-metallic with large indirect gaps of $\sim$3~eV along
M-$\Gamma$ symmetry line (E$_{\rm g}^{\rm {M-\Sigma}}$) and 2.3~eV
at W symmetry point (E$_{\rm g}^{\rm {X-W}}$), respectively. The
nonmetallic nature of ScN in ZB phase was also obtained by an
indirect gap of 2.36~eV at W point (E$_{\rm g}^{\rm {X-W}}$) by
LDA scheme~\cite{11}. These GGA and LDA calculations of ScN in ZB
phase~\cite{9,11}, have given the direct gap of 2.4 and 2.42~eV at
X symmetry point, respectively. In the literature, ScN in B2 phase
was reported to be metallic by DFT calculations within
GGA~\cite{9,10} and LDA~\cite{11} schemes.

In the literature, although ScN has not been worked so far, hybrid
FP-LAPW calculations within the framework of DFT and different
exchange-correlation functionals have given accurate electronic
features for nitride compounds and alloys~\cite{26,27} comparable
with the corresponding measured ones due to the possible strong
hybridization between the $2p$ orbitals of N and the corresponding
cationic states. In addition, the orthogonalized norm-conserving
pseudopotential (NCPP) method in which the plane wave basis
functions are orthogonalized to core-like orbitals has been
reported to be a very promising method in electronic band
structure calculations of nitride alloys~\cite{28} for describing
the experimental optical data, together with the FP-LAPW method
within virtual crystal approximation~\cite{28}.

In the present work, we have examined the structural and
electronic properties of ScN in stable (RS) and meta-stable phases
(CsCl, ZB, WZ) by DFT calculations mainly within two GGA schemes
which have not been used before for ScN. We have aimed to
introduce a comparative study for the structural and electronic
features of ScN such as the lattice constant, bulk modulus,
cohesive energy, energy gaps and the effective masses  of
electrons and holes. The present work has also comprised the
electronic band structure of stable ScN corrected by an on-site
Coulomb self- and exchange-correlation potential approximation
(U$^{\rm SIC}$)~\cite{29}.

\section{Method of calculations}
The present DFT calculations on the structural and electronic
properties of ScN compound have been performed using FP-LAPW
method implemented in WIEN2k code~\cite{30}. In the literature,
the exchange-correlation energy of DFT has been defined by local
density approximation (LDA)~\cite{16} for the systems having
uniform electron charge density. But for the systems of
non-uniform charge density, the exchange-correlation energy of LDA
has been corrected by gradient of the charge density within
different generalized gradient approximations (GGA). In the
present total energy and electronic band structure calculations of
ScN in B1-B4 phases, four different approximations of
exchange-correlation energies have been considered. In one of the
approximations, exchange and correlation energies have been
defined simply by LDA~\cite{16}, without regarding the homogeneity
of the real charge density. In the second approximation, exchange
and correlation energies of LDA have been corrected by GGA of
Perdew-Burke-Ernzerhof (PBE)~\cite{19}. The generalized gradient
functional of Perdew-Burke-Ernzerhof~\cite{19} has retained
correct features of LDA~\cite{16} and satisfied only those which
are energetically significant. In the third approximation, GGA of
Engel-Vosko (EV)~\cite{31} and GGA of Perdew and Wang
(PW)~\cite{32} have been used to correct the exchange and
correlation energies, respectively. Since the generalized gradient
functional of Engel-Vosko~\cite{31} was designed to give a better
exchange potential ($V_{\rm x}$) only, the standard correlation
potential of LDA~\cite{16} in the third approximation has been
corrected by another functional, namely, GGA of Perdew and
Wang~\cite{32}. The functional of Perdew and Wang~\cite{32} has
incorporated some inhomogeneity effects while retaining many of
the best features of the local density approximation. In the last
approximation, the exchange energy of LDA~\cite{16} was corrected
by GGA of Engel-Vosko~\cite{31}, but the correlation energy was
defined directly by LDA~\cite{16}. The exchange-correlation energy
approaches considered in this work have been labeled as LDA,
PBE-GGA, EV$_{\rm ex}$-PW$_{\rm co}$-GGA and EV$_{\rm
ex}$-GGA-LDA$_{\rm co}$. The acronyms have been produced either
based on the key word of the approach (LDA) or on the name of the
authors (PBE-GGA, EV$_{\rm ex}$-PW$_{\rm co}$-GGA, EV$_{\rm
ex}$-GGA-LDA$_{\rm co}$) who developed the corresponding exchange
and correlation functionals. Here, the subscripts of exchange (ex)
and correlation (co) functionals are exclusively used for the
cases where the exchange and correlation functionals are
different. It has been considered that EV$_{\rm ex}$-PW$_{\rm
co}$-GGA and EV$_{\rm ex}$-GGA-LDA$_{\rm co}$ schemes can provide
significant improvement for the structural and electronic
properties of ScN, respectively. In the literature, it was
reported that LDA+U and GGA+U schemes can also improve the band
gap energies of transition metal compounds and alloys by
reproducing quite well the localized nature of the $d$-electrons (or
$f$-electrons)~\cite{33,34,35}. Since the ab-initio calculations are
difficult to perform, the strong correlations like in transition
metal compounds and alloys are often based on a model Hamiltonian
approach in which the important parameter of U improves the
effective Coulomb interactions between the localized $d$-electrons.
In the present work, the electronic band structure of RS-ScN
calculated by EV$_{\rm ex}$-PW$_{\rm co}$-GGA has been improved by
U$^{\rm SIC}$ method~\cite{29,36} introduced in WIEN2k
code~\cite{30}. The present EV$_{\rm ex}$-PW$_{\rm
co}$-GGA+U$^{\rm SIC}$ scheme has rectified on-site Coulomb self-
and exchange-correlation interactions of the localized $d$-orbitals
of Sc by the potential parameter of U~\cite{29,36}. In Anisimov et
al's paper~\cite{29}, the meaning of the U parameter were defined
as the cost Coulomb energy for the placement of two $d$-electrons on
the same site.  The Coulomb interaction was defined to be
$({1}/{2})$U$\sum_{i\not=j}$n$_{i}$n$_{j}$~\cite{37} for
$d$-orbitals. Here, n$_{i}$ are $d$-orbital occupancies. The Coulomb
interaction term included into the total energy functional of
EV$_{\rm ex}$-PW$_{\rm co}$-GGA  has given the orbital energies of
ScN  as $\xi_{i}$=$\xi$$_{({\rm EV}_{\rm ex}}$$_{-{\rm PW}_{\rm
co}}$$_{-{\rm GGA})}$+U(${1}/{2}-n_{i})$~\cite{37}. The
shifting of the corresponding orbital energies in EV$_{\rm
ex}$-PW$_{\rm co}$-GGA+U$^{\rm SIC}$ calculations gives a
qualitative improvement for the energy gap of RS-ScN. The present
Coulomb interaction parameter of U is calculated to be 4.08~eV in
WIEN2k~\cite{30} to have the maximum approach to the measured
indirect band gap of RS-ScN.

In the present work, ScN has been studied in RS, CsCl, ZB, and WZ
structures. The unit cells of RS, CsCl, and ZB consist of two
basis atoms; Sc at $(0,\, 0,\, 0)$ and N at $(0.5a,\, 0.5a,\,
0.5a)$ in fcc structure, Sc at $(0,\, 0,\, 0)$ and N at $(0.5a,\,
0.5a,\, 0.5a)$ in bcc structure, and Sc at $(0,\, 0,\, 0)$ and N
at $(0.25a,\, 0.25a,\, 0.25a)$ in fcc structure, respectively,
where $a$ is the lattice constant parameter. The WZ structure with
space group of $F6_{3}mc$ has four atoms in the unit cell; Sc
atoms at $(a/3,\, 2a/3,\, 0)$ and $(2a/3,\, a/3,\, 0.5c)$, N atoms
at $(a/3,\, 2a/3,\, u*c)$ and $(2a/3,\, a/3,\, (0.5+u)*c)$, where
$a$ and $c$ are the periods in x-y plane and along z direction,
respectively. The z directional distance, $u$, is defined  between
the layers of Sc and N atoms. In FP-LAPW calculations, each unit
cell is partitioned into non-overlapping muffin-tin spheres around
the atomic sites. Basis functions are expanded in combinations of
spherical harmonic functions inside the non-overlapping spheres.
In the interstitial region, a plane wave basis is used and
expansion is limited with a cutoff parameter, $R_{\rm MT}K_{\rm
MAX}$=7. Here, $R_{\rm MT}$ is the smallest radius of the sphere
in the unit cell, $K_{\rm MAX}$ is the magnitude of the largest K
vector used in the plane wave expansion. The muffin-tin radius is
adopted to be 1.8 and 1.67~a.u. for Sc and N atoms, respectively.
In the calculations, the electrons of Sc and N atoms in
$3s^{2}3p^{6}4s^{2}3d^{1}$ and $2s^{2}2p^{5}$ shells respectively,
are treated as valence electrons by choosing a cutoff energy of
--6.0~Ry. The core states are treated within the spherical part of
the potential only and are assumed to have a spherically symmetric
charge density totally confined inside the muffin-tin spheres. The
expansion of spherical harmonic functions inside the muffin-tin
spheres is truncated at $\it{l}$=10. The cutoff for Fourier
expansion of the charge density and potential in the interstitial
region is fixed to be $G_{\rm MAX}=16 \sqrt{\rm Ry}$. The FP-LAPW
parameters presented in this work have been obtained after a few
trials around their fixed values.

\begin{figure}[ht]
\centerline{\includegraphics[width=12cm]{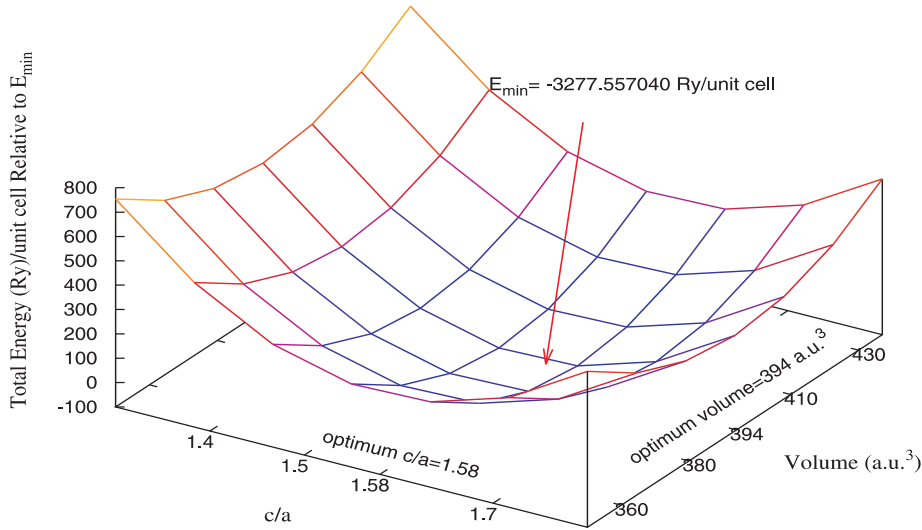}}
\caption{(Color on-line) The relative total energy
(per unit cell) versus volume and $c/a$ within EV$_{\rm
ex}$-PW$_{\rm co}$-GGA for WZ-ScN.} \label{figure 1}
\end{figure}

The ScN in RS, CsCl and ZB structures have been optimized with
respect to the volume of the unit cells by minimizing the total
energy. The equilibrium lattice constants of ScN in RS, CsCl, and
ZB phases are determined by fitting the total energies to the
Murnaghan's equation of state~\cite{38}. The equilibrium structure
of ScN in WZ phase that corresponds to the minimum total energy
has been obtained by the application of both volume and geometry
optimizations. The volume optimization used for all structures is
provided with the energy criterion of 0.01~mRy. The optimum volume
for WZ phase corresponds to the optimum $c/a$ ratio and $a$ value
has been found by fitting the total energies to a quadratic
function in a least square fitting method. The z directional
distance $u$ between the Sc and N layers in WZ phase has been
obtained by geometry optimization at the optimum volume of the
unit cell. The geometry optimization forces the atoms in the unit
cell to move towards their equilibrium positions. In the geometry
optimization, all forces on the atoms are converged to less than
1~mRy/a.u. The variation of total energy with respect to the
volume and $c/a$ ratio of WZ structure is plotted in figure~\ref{figure 1}. The
present structural and electronic band calculations have been
performed using 21x21x21 grids and correspond to 1000 k points
sufficiently defined in the irreducible wedge of the Brillouin
zone for ScN in B1-B4 phases.

In the present work, the cohesive energies (energy/atom-pair) of ScN
structures (B1-B4) have been calculated by
\begin{eqnarray}
E_{\rm coh.}=E_{\rm ScN}-ME_{\rm N}-ME_{\rm Sc}\,.
\end{eqnarray}
Where, E$_{\rm N}$ and E$_{\rm Sc}$ are the values of the
self-atomic energies of N and Sc atoms. E$_{\rm ScN}$ is the
minimum of the total energy per unit cell that corresponds to the
equilibrium structures of the ScN phases. Here, M defines the
number of N and Sc atoms in the unit cell of the phases. The
self-atomic energies of either atoms are calculated accurately in
a fcc super cell with a lattice constant of 25~a.u.

\section{Results and discussion}

\subsection{Structural properties}

\begin{figure}[h]
\centerline{\includegraphics[width=6cm]{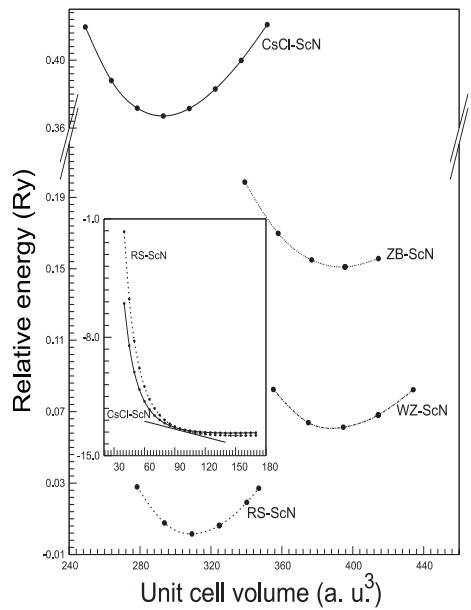}}   
\caption{The relative total energy (per unit cell) versus volume
within EV$_{\rm ex}$-PW$_{\rm co}$-GGA for ScN in RS, WZ, ZB, and
CsCl structures. The inset figure displays the cross point on the
variation of the total energies (per unit cell) of the RS-ScN
(dotted line) and CsCl-ScN (solid line) structures as a function
of volume.} \label{figure 2}
\end{figure}

The total energies (per unit cell) of  RS, CsCl, WZ and ZB
structures of ScN calculated within EV$_{\rm ex}$-PW$_{\rm
co}$-GGA scheme are plotted as a function of the volume of the
structures in figure~\ref{figure 2}. The curvature of variations is made clear
by plotting the total energies (per unit cell) of the structures
relative to the minimum total energy of the RS-ScN. The minimum of
the total energies of the phases with
E(RS-ScN)$<$E(WZ-ScN)$<$E(ZB-ScN)$<$E(CsCl-ScN) rank indicates
that the RS is the most stable structure of ScN as it was reported
in the other works~\cite{1,3,4,5,6,7,8,9,10,11}. The cohesive
energies of ScN are calculated to be --12.40, --13.87, --14.06 and
--14.48~eV for B2, B3, B4 type metastable and B1 type most stable
structures, respectively. In the cohesive energy
calculations~[equation~(2.1)], the atomic self-energies of Sc and
N atoms are calculated to be --1528.609 and --109.136~Ry,
respectively. The present cohesive energies are found to be
different ($\sim$1~eV) from the ones~\cite{9,10} defined by only
the minimum total energies (per unit cell) of the structures.

\begin{table}[h]
\caption{The optimized lattice constant [a(\AA)], bulk modulus
[B(GPa)], and first-order pressure derivative of bulk modulus
[B$^{'}$] calculated by LDA (I), PBE-GGA (II), EV$_{\rm
ex}$-PW$_{\rm co}$-GGA (III), and EV$_{\rm ex}$-GGA-LDA$_{\rm co}$
(IV) approaches for ScN in RS, CsCl, ZB, and WZ phases.}\label{tab1}
$$
\begin{array}{|l|l|llll|lll|}
\hline
&          &              & &                              &   & &&\\
& & $Present$&\hspace{-2mm}$Work$ && &$Other$~$Works$ & & \\

&         &$I$    &\hspace{-2mm}$II$         &$III$ &$IV$        &$Theor.$&&$Exp.$\\
\hline
          &       &      &     &             &      &                             & & \\
          &  a  &4.432 &\hspace{-2mm}4.513 &\hspace{-3mm}4.510 &4.755 &4.44^{9}, 4.54^{9}    & &4.501^{1} \\
          &       &      &                   &      &      &4.520^{10}, 4.651^{11}& &4.5^{3-8}   \\
          &       &      &                   &      &      &4.42^{12}, 4.51^{13}  & &            \\
          &       &      &                   &      &      &4.50^{12,14,15}       & &   \\
          &       &      &                   &      &      &4.455^{14}, 4.533^{14}& &  \\
$RS-ScN$  &$B$    &230.44&\hspace{-2mm}195.95&202.35&137.35&220^{9}, 201^{9,12}   & &182\mp{40}^{1} \\
          &       &      &                   &      &      &201.576^{10}          & &         \\
          &       &      &                   &      &      &210.364^{11}, 235^{12}& &   \\
          &       &      &                   &      &      &197^{13}, 221^{14}    & & \\
          &       &      &                   &      &      &196^{14}              & & \\
          &$B$^{'}&4.41  &\hspace{-2mm}4.31  & 4.25 & 4.16 &3.31^{9}, 3.89^{10,13}& & \\
          &       &      &                   &      &      &3.12^{11}, 4.27^{14}  & &  \\
          &       &      &                   &      &      &4.36^{14}             & & \\
\hline
          &       &      &                   &      &      &                      & &  \\
          &  a  & 2.727&\hspace{-2mm}2.787 &2.788 &2.961 &2.81^{9}, 2.79^{10}, 2.926^{11}& & \\
$CsCl-ScN$&  $B$  &212   &\hspace{-2mm}182   &180   &118   &170^{9}, 178.60^{10}, 159.759^{11}& & \\
          &$B$^{'}&4.04  &\hspace{-2mm}4.28  &4.15  & 3.93 &3.47^{9}, 4.43^{10}, 3.034^{11}& & \\
\hline
          &       &      &                   &      &      &                               & &  \\
          &  a  &4.804 &\hspace{-2mm}4.893 &4.888 &5.142 &4.88^{9}, 4.939^{11}      & & \\
$ZB-ScN$  &$B$    &159.12&\hspace{-2mm}141.18&142.35&89.83 &153^{9}, 145.195^{11}     & &\\
          &$B$^{'}&3.85  &\hspace{-2mm}3.79  &3.77  &5.06  &3.34^{9}, 3.43^{11}       & & \\
\hline
          &       &      &                   &      &      &                          &  &  \\
          &a    &3.424 &\hspace{-2mm}3.497 &3.494 &3.680 &3.49^{9}, 3.45^{11}       &  &\\
          &c/a  &1.578 &\hspace{-2mm}1.582 &1.581 &1.581 &1.6^{9}, 1.552^{11}  & &\\
$WZ-ScN$  &u    &0.389&\hspace{-2mm}0.389&0.389&0.387&0.38^{9}, 0.377^{11} &  &\\
          &$B$    &146   &\hspace{-2mm}131   &131   &86    &156^{9}, 132.91^{11}& & \\
          &$B$^{'}&4.18  &\hspace{-2mm}3.93  &3.43  &3.88  &2.16^{9}, 3.29^{11}& &\\
\hline
\end{array}
$$
\end{table}


The equilibrium lattice constants, bulk moduli, and first order
pressure derivative of the \linebreak bulk moduli of ScN in B1-B4 phases
calculated within LDA, PBE-GGA, EV$_{\rm ex}$-PW$_{\rm co}$-GGA,
and EV$_{\rm ex}$-GGA-LDA$_{\rm co}$ schemes are tabulated in
table~\ref{tab1}, together with the available measured and calculated
values of the other groups, for comparison. The equilibrium
lattice constants of 4.510 and 4.513~\AA~obtained for RS-ScN
structure by EV$_{\rm ex}$-PW$_{\rm co}$-GGA and PBE-GGA
calculations, respectively, are found to be very close to the
measured values of 4.501~\cite{1} and 4.5~\AA~\cite{3,4,5,6,7,8}.
On the other hand, the equilibrium lattice constant of RS-ScN
obtained by the present EV$_{\rm ex}$-PW$_{\rm co}$-GGA and
PBE-GGA calculations are found to be very close to the values
calculated by FP-LAPW and ab-initio PP methods within PBE-GGA and
PCVJPSF
(Perdew-Chevary-Vosko-Jackson-Pederson-Singh-Fiolhais)-GGA~\cite{39}
schemes~\cite{10,12,13}. The present bulk moduli of 202.35 and
195.95~GPa found for RS-ScN within EV$_{\rm ex}$-PW$_{\rm co}$-GGA
and PBE-GGA's, respectively, are the best ones providing the
measured value of 182$\mp$40~GPa~\cite{1}. The present bulk moduli
of EV$_{\rm ex}$-PW$_{\rm co}$-GGA and PBE-GGA's are also found to
be in the range of 196--202~GPa defined by the corresponding results
of ab-initio PP and FP-LAPW calculations within PBE-GGA and
PCVJPSF-GGA schemes~\cite{13,14,9,10,12}. The equilibrium lattice
constants and bulk moduli of RS-ScN calculated by LDA and EV$_{\rm
ex}$-GGA-LDA$_{\rm co}$'s are either underestimated or
overestimated with respect to the measured ones given in table~\ref{tab1}.
The values of the first order pressure derivative of the bulk
moduli (table~\ref{tab1}) presented for RS-ScN are all close to each other
and in agreement with the results of {ab-initio} PP calculations
of LDA and PBE-GGA's~\cite{14}. As it is determined for ScN in RS
phase, the structural features of ScN in B2-B4 phases calculated
by EV$_{\rm ex}$-PW$_{\rm co}$-GGA and PBE-GGA's are very close to
each other (table~\ref{tab1}). The present equilibrium lattice constants,
bulk moduli, and first order pressure derivatives of the bulk
moduli of ScN in B2-B4 phases could not be compared, because of
the lack of the measured ones in the literature. But, they are
(especially the results of LDA, PBE-GGA and EV$_{\rm ex}$-PW$_{\rm
co}$-GGA) comparable with the results of the other groups
calculated within LDA~\cite{9,11} and PBE-GGA~\cite{9,10} schemes.

\begin{figure}
\centerline{\includegraphics[width=7cm]{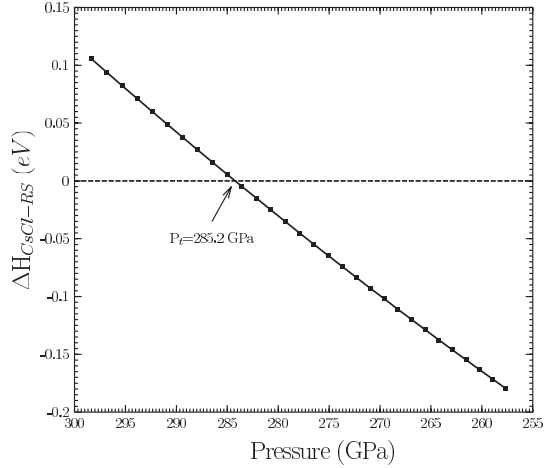}}
\caption{The difference of enthalpy between the
CsCl and RS phases ($\Delta$H$_{\rm CsCl-RS}$) versus pressure.}
\label{figure 3}
\end{figure}
The total energies (per unit cell) of RS-ScN and CsCl-ScN relative
to a reference total energy are plotted  in the small frame of
figure~\ref{figure 2} for low unit cell volumes of the structures. The crossing
of the total energy curves observed in the inset figure indicates
the phase transition from RS to CsCl structure at high pressure.
The enthalpy of the RS phase equals to that of the CsCl phase at
the cross point. Therefore, the necessary pressure providing the
transition from RS to CsCl corresponds to the zero difference
between the enthalpy of the structures. The enthalpy of the RS and
CsCl phases is evaluated by Gibb's free energy
\begin{eqnarray}
H=E_{\rm tot}+PV-TS
\end{eqnarray}
at T=0~K. In figure~\ref{figure 3}, the difference between the enthalpy of the
CsCl and RS structures ($\Delta H_{\rm CsCl-RS}$) decreases  to
zero at the transition pressure of 285.2~GPa. The present
transition pressure is found to be smaller than the ones
(341~GPa~\cite{9}, 332.75~GPa~\cite{11}) calculated directly from
the slope of the common tangent.

\subsection{Electronic properties}
\begin{figure}[!h]
\centerline{\includegraphics[width=7cm,angle=-90]{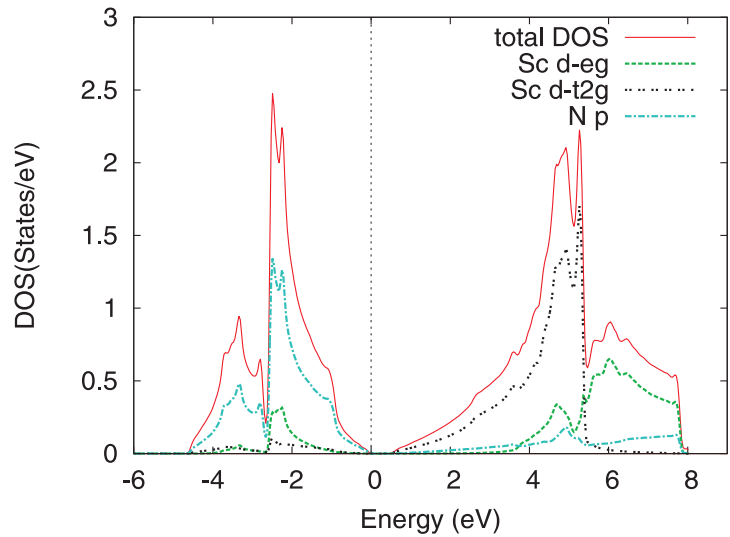}} 
\caption{(Color on-line) The total and partial DOS within EV$_{\rm
ex}$-PW$_{\rm co}$-GGA for RS-ScN.} \label{figure 4}
\end{figure}

%
\begin{table}[h]
\caption{The calculated and experimental energy band gaps [E$_{\rm
g}$(eV)] of RS-ScN.}\label{tab2}
$$
\begin{array}{|l|rrrrrl|}
\hline
                              &                        &              &             &                  &                  &  \\
{\rm Approach}                      &$E$_{\rm g}^{\Gamma-\Gamma} &$E$_{\rm g}^{\rm {X-X}} &$E$_{\rm g}^{\rm W-W}&~~$E$_{\rm g}^{\rm {\Gamma-X}}&~~$E$_{\rm g}^{\rm\Gamma-K}&$~~~~E$_{\rm g}^{\rm {X-W}} \\
\hline
                              &                        &              &             &                   &                  &  \\
$LDA$^{a}                     &2.40                    &0.68          &6.43         &~~-0.24            &~~3.69              &~~~~4.61   \\
   $PBE-GGA$^{a}              &2.43                    &0.91          &6.22         &~~ -0.01            &~~3.59              &~~~~4.52  \\
   $EV$_{\rm ex}$-PW$_{\rm co}$-GGA$^a              &2.60                    &1.33          &6.17         &~~0.46              &~~3.73              &~~~~4.60  \\
   $EV$_{\rm ex}$-PW$_{\rm co}$-GGA+U$^a            &3.28                    &1.82          &6.81         &~~0.90              &~~4.33             &~~~~5.25 \\
   $EV$_{\rm ex}$-GGA-LDA$_{\rm co}$$^a                 &2.29                    &1.46          &5.37         &~~0.62              &~~3.19              &~~~~4.03   \\
   $GGA$^9                    &2.40                    &0.90          &6.40         &~~0.0               &~~2.30              &  \\
   $GGA$^{10}                 &                        &0.9           &             &~~\sim0.0           &                  & \\
   $LDA$^{11}                 &                        &              &             &~~\sim0.0           &                  & \\
   $LDA$^{12}                 &                        &              &             &~~0.0               &                  &  \\
   $sX-LDA$^{12}              &                        &2.41          &             &~~1.58              &                  &   \\
   $GGA$^{13}                 &2.41                    &0.90          &6.21         &~~0.02              &~~2.22              &   \\
  $LDA$^{14}                  &2.34                    &0.75          &             &~~-0.15             &                  & \\
  $GGA$^{14}                  &2.43                    &0.87          &             &~~-0.03             &                  & \\
  $OEPx(cLDA)$-$G$_{o}$W$_{o}^{14} &3.51               &1.98          &             &~~0.84              &                  &  \\
  $LDA$-$G$_{o}$W$_{o}^{14}        &3.71               &2.06          &             &~~1.14              &                  &  \\
  $OEPx(cLDA)$^{14}                &4.53               &2.59          &             &~~1.70              &                  & \\
  $GGA$^{15}                       &                   &              &             &~~0.54              &                  & \\
  $EXX-GGA$^{20}                   &4.70               &2.90          &             &~~1.60              &                  &  \\
  $LDA$-$GW$^{21}                  &4.3\phantom{o}     &2.0           &             &~~0.9               &                  & \\
  {\rm exp.}^{3}                         &                   &1.8-2.1       &             &                  &                  &  \\
  {\rm exp.}^4                           &                   &2.1           &             &                  &                  &  \\
  {\rm exp.}^7                           &2.1-2.4            &2.15          &             &~~0.9\mp0.1         &                  &  \\
  {\rm exp.}^8                           &>2.6              &               &             &~~2.4               &                  & \\
  {\rm exp.}^{20}                        &\sim3.8           &2.4\mp0.3      &             &~~1.3\mp0.3         &                  &  \\
\hline
\end{array}
$$
\hspace{5mm}$^{a}$ present~ work
\end{table}
%
The electronic band structures of ScN in RS, CsCl, ZB, and WZ
phases have been calculated along  the various symmetry lines
within LDA, PBE-GGA, EV$_{\rm ex}$-PW$_{\rm co}$-GGA, and EV$_{\rm
ex}$-GGA-LDA$_{\rm co}$ schemes. The energy gaps of the RS-ScN
correspond to different symmetry directions, and the points are
tabulated in table~\ref{tab2}, together with the available measured and
calculated energy gaps of the other groups, for comparison. The
electronic band structure calculations of RS-ScN phase have given
the top of the valence band at $\Gamma$ point. The bonding
analysis given in~\cite{40} has shown that RS-AlN has three $p$-like
bonding, three $d$-like antibonding $t_{\rm 2g}$, and two $d$-like
nonbonding $e_{\rm g}$ bands formed by the hybridization of
three-valence $p$ states of N with the five $d$ states of Sc.
According to the present partial density of states (DOS)
calculations plotted in figure~\ref{figure 4}, the upper valence bands are
formed mainly by the N $p$-orbitals with some mixture of Sc
$d$-orbitals, while the conduction bands are predominantly
originated from Sc $d$-$t_{\rm 2g}$ states with some admixture of
the Sc $d$-$e_{\rm g}$ and N $p$-states.
The minimum of the conduction
band is mainly of Sc $3d$ character. The present identification for
the conduction and valance bands of ScN is consistent with the
partial density of states analysis given in~\cite{14}. The present
LDA and PBE-GGA calculations have given the negative indirect gap
at X point for RS-ScN. The negative or approximately zero indirect
gap at X point was also reported in~\cite{9,10,11,12,13,14} for
RS-ScN by the same approximations. The indirect energy gap of
RS-ScN is 0.46 and 0.62~eV by the present calculations of EV$_{\rm
ex}$-PW$_{\rm co}$-GGA and EV$_{\rm ex}$-GGA-LDA$_{\rm co}$'s,
respectively. Therefore, EV$_{\rm ex}$-PW$_{\rm co}$-GGA and
EV$_{\rm ex}$-GGA-LDA$_{\rm co}$'s are found to be more accurate
than PBE-GGA to produce a positive indirect gap for ScN in RS
phase. These indirect gap values obtained by the present
non-corrected EV$_{\rm ex}$-PW$_{\rm co}$-GGA and EV$_{\rm
ex}$-GGA-LDA$_{\rm co}$ calculations are in agreement with the
value of 0.54~eV reported by non-corrected PP-GGA
calculations~\cite{15}. In the literature, different correction
approximations used in LDA and GGA schemes were found to be
effective to produce a larger indirect gap close to the measured
value of $\sim$1~eV for RS-ScN~\cite{12,14,20,21}. The present
EV$_{\rm ex}$-PW$_{\rm co}$-GGA+U$^{\rm SIC}$ calculations have
also improved the highest valence and lowest conduction band
states  of RS-ScN by giving an indirect gap of 0.9~eV. It is found
that the corrected Coulomb interaction term by the U parameter of
4.08~eV pushes the N $2p$ bands down and Sc $3d$-$t_{\rm 2g}$ bands up
in the present EV$_{\rm ex}$-PW$_{\rm co}$-GGA+U$^{\rm SIC}$
calculations to enlarge the indirect energy gap by an amount of
0.44~eV. Since the valence and conduction bands in RS-ScN
structure are formed by the hybridization of $d$-orbitals of Sc and
$p$-orbitals of N atoms~\cite{40}, the correction on the Coulomb
self- and exchange-correlation interactions of the localized
$d$-orbitals of Sc in EV$_{\rm ex}$-PW$_{\rm co}$-GGA+U$^{\rm SIC}$
calculations has indirectly improved N $2p$ bands. In the same
correction method, the direct band gap of RS phase at X point is
calculated to be 1.82~eV with respect to the measured and
calculated values in the ranges of 1.8--2.4~eV~\cite{3,4,7,20} and
1.98--2.90~eV~\cite{12,14,20,21}, respectively. The present
$\Gamma$ point direct energy gap of 3.28~eV for B1 phase is found
to be comparable with the results of optical transmission~\cite{8}
and reflection~\cite{20} measurements and OEPx(cLDA)-G$_{\rm
o}$W$_{\rm o}$~\cite{14} and LDA-G$_{\rm o}$W$_{\rm o}$~\cite{14}
calculations (table~\ref{tab2}). The present electronic band structure of
RS-ScN within EV$_{\rm ex}$-PW$_{\rm co}$-GGA scheme is plotted in
figure~\ref{figure 5}. The improvement of the band structure due to EV$_{\rm
ex}$-PW$_{\rm co}$-GGA+U$^{\rm SIC}$ calculations is also shown in
the same plot.
\begin{figure}[h]
\centerline{\includegraphics[width=7cm]{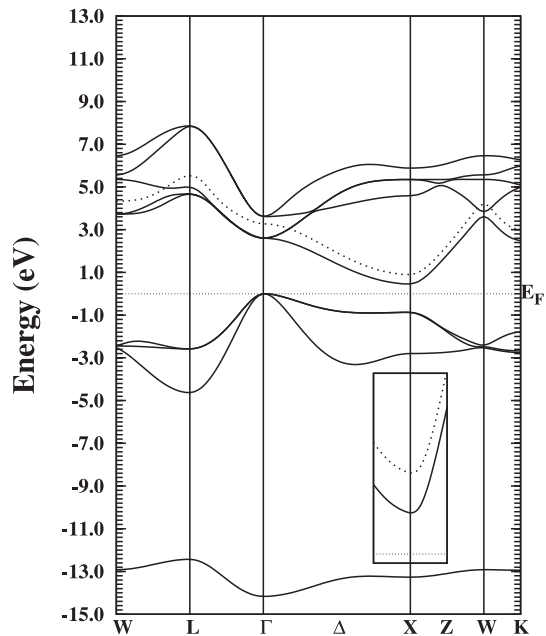}}
\caption{Electronic band structure of RS-ScN within
EV$_{\rm ex}$-PW$_{\rm co}$-GGA. The dotted line shows the
correction of the conduction band state within EV$_{\rm
ex}$-PW$_{\rm co}$-GGA+U$^{\rm SIC}$ calculations.} \label{figure
5}
\end{figure}

In this work, the effective electron and hole masses (heavy and
light) are calculated using the electronic band structure of
RS-ScN obtained by EV$_{\rm ex}$-PW$_{\rm co}$-GGA+U$^{\rm SIC}$
scheme (figure~\ref{figure 5}). The conduction band and valance band effective
masses at the X and $\Gamma$ points, respectively, are calculated
by fitting a quadratic function to the corresponding band
structure energies. The closely spaced k points in a very small
range around the X point have yielded the electron effective mass
of 0.22\,$m_{e}$\,. The electron effective mass in RS-ScN was
determined between 0.1 and 0.2\,$m_{e}$ by extrapolating the
measured values of infrared reflectivity for high carrier
concentrations to low ones~\cite{41}. Since the edge of the
conduction band along $\Delta (\Gamma$-X) and Z (W-X) directions
approaches the X point with quiet different slopes (inset plot in
figure~\ref{figure 5}), the electron effective mass is also calculated along
$\Delta$ and Z directions, separately. The electron effective
masses are calculated to be 1.621\,$m_{e}$ and 0.223\,$m_{e}$
along $\Delta $ and Z directions around X point with respect to
the ones reported in the range of 1.441--1.625\,$m_{e}$ and
0.124--0.253\,$m_{e}$\,, respectively, by LDA~\cite{7,14},
GGA~\cite{14}, OEPx(cLDA)~\cite{14}, and OEPx(cLDA)-G$_{\rm
o}$W$_{\rm o}$~\cite{14} schemes. In the present work, the
effective mass for the heavy and light holes are calculated to be
0.93 and 0.19\,$m_{e}$ at $\Gamma$ point, respectively.

\begin{figure}[ht]
\includegraphics[width=7cm]{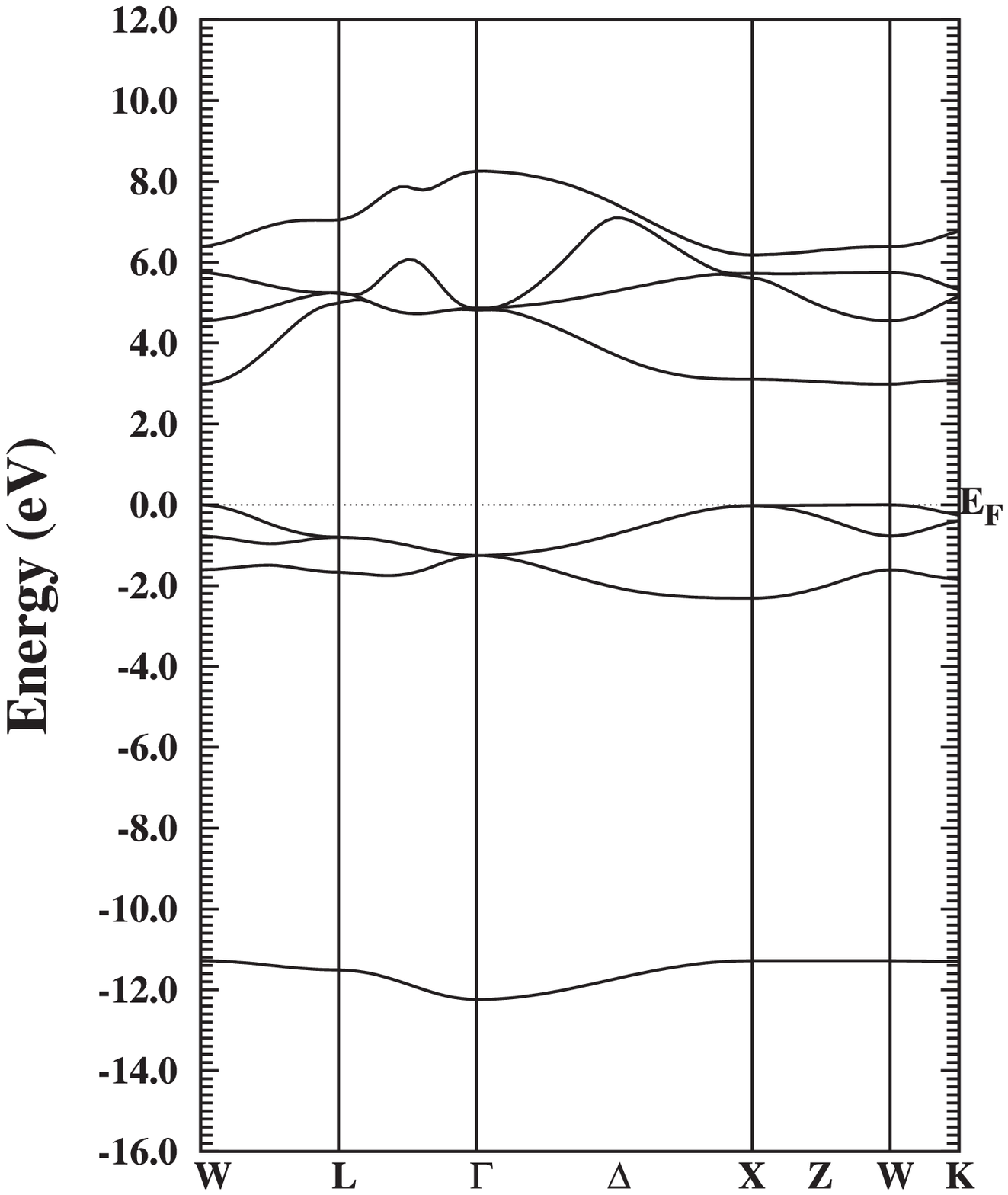}%
\hfill%
\includegraphics[width=7cm]{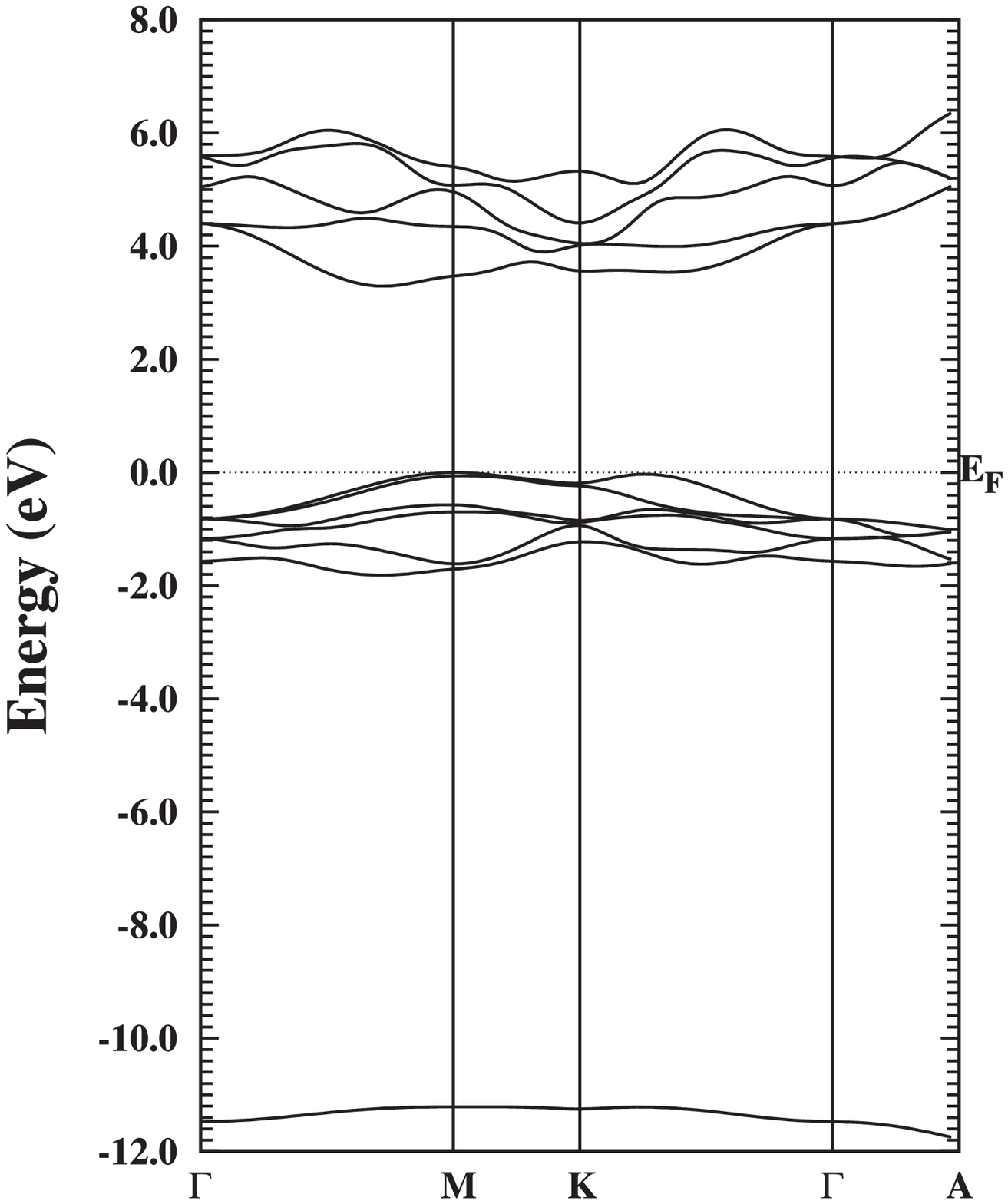}%
\\%
\parbox[t]{0.48\textwidth}{%
\caption{Electronic band structure of ZB-ScN within
EV$_{\rm ex}$-PW$_{\rm co}$-GGA.} \label{figure 6}%
}%
\hfill%
\parbox[t]{0.48\textwidth}{%
\caption{Electronic band structure of WZ-ScN within
EV$_{\rm ex}$-PW$_{\rm co}$-GGA.} \label{figure 7}%
}%
\end{figure}
%
\begin{figure}[!h]
\centerline{\includegraphics[width=7cm]{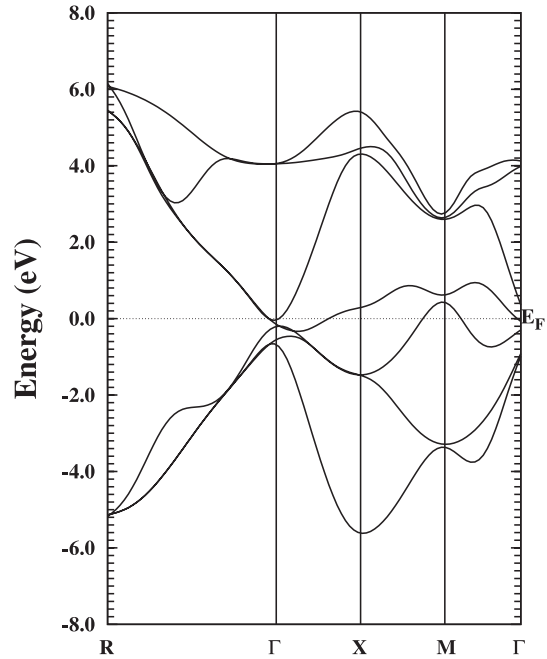}}
 \caption{Electronic band structure of CsCl-ScN
within EV$_{\rm ex}$-PW$_{\rm co}$-GGA.} \label{figure 8}
\end{figure}
The electronic band structures of ScN in B2-B4 metastable phases
have been also calculated by LDA, PBE-GGA, EV$_{\rm ex}$-PW$_{\rm
co}$-GGA, and EV$_{\rm ex}$-GGA-LDA$_{\rm co}$ schemes. The band
structures of the ZB, WZ, and CsCl phases calculated by EV$_{\rm
ex}$-PW$_{\rm co}$-GGA are plotted in figures~\ref{figure 6}--\ref{figure 8}. The energy
gaps of the ZB and WZ metastable structures correspond to
different symmetry lines, and the points are tabulated in table~\ref{tab3}.
The calculations have yielded non-metallic band structures for
ZB-ScN and WZ-ScN with the top of the valence band at W and M
symmetry points, respectively. The ZB-ScN structure is found to be
a direct band gap material even if the direct (E$_{\rm g}^{\rm
W-W}$) and indirect (E$_{\rm g}^{\rm W-X}$) band gaps are very
close to each other. However, ZB-ScN was an indirect band gap
(E$_{\rm g}^{\rm {X-W}}$) material in~\cite{9,11}. In
reference~\cite{9}, the ZB-ScN structure was a direct band gap
(E$_{\rm g}^{\rm W-W}$) material when the electronic band
structures were calculated  with the lattice constant of RS-ScN.
The present direct band gap energies of PBE-GGA (2.44~eV) and LDA
(2.30~eV) schemes for ZB structure are found to be comparable with
the corresponding values of 2.4 and 2.36~eV given by similar
approximations~\cite{9,11}. The present direct band gap of ZB
phase at W point is enlarged to $\sim$3~eV by EV$_{\rm
ex}$-PW$_{\rm co}$-GGA (3~eV) and EV$_{\rm ex}$-GGA-LDA$_{\rm co}$
(2.82~eV) calculations. The indirect band gap of WZ phase is
calculated to be $\sim$3~eV along M-$\Sigma$ direction by LDA and
PBE-GGA schemes. The large indirect band gap of ScN in WZ
structure was also reported in~\cite{9}. As it is found for ZB
structure, the indirect band gap of WZ-ScN is enlarged by EV$_{\rm
ex}$-PW$_{\rm co}$-GGA (3.29~eV) and EV$_{\rm ex}$-GGA-LDA$_{\rm
co}$ (3.08~eV)'s. According to all approximations considered in
this work, ScN in CsCl phase has a metallic band structure
(figure~\ref{figure 8}) as it was reported earlier by LDA and GGA
calculations~\cite{9,11}; the conduction and valance bands are
observed to be mixed completely. Because of the lack of the
measurements on metastable structures of ScN, the present results
are compared only by the calculated ones.

\begin{table}
\caption{The energy band gaps [E$_{\rm g}$(eV)] of ZB-ScN and
WZ-ScN phases calculated by LDA (I), PBE-GGA (II), EV$_{\rm
ex}$-PW$_{\rm co}$-GGA (III), and EV$_{\rm ex}$-GGA-LDA$_{\rm co}$
(IV) approaches.}\label{tab3}
$$
\begin{array}{|l|l|llllllllll|}
\hline
& & & & & & & & & & & \\
\rotatebox{90}{\small Phase}&\rotatebox{90}{\small
Approach}&$E$_{\rm g}^{\Gamma-\Gamma} &$E$_{\rm g}^{\rm {X-X}}
&$E$_{\rm g}^{\rm W-W}&$E$_{\rm g}^{\rm L-L}&$E$_{\rm g}^{\rm
{\Gamma-X}}&$E$_{\rm g}^{\rm \Gamma-W}&$E$_{\rm g}^{\rm
\Gamma-L}&$E$_{\rm g}^{\rm W-\Gamma}&$E$_{\rm g}^{\rm
W-L}&$E$_{\rm g}^{\rm W-X}
\\
\vspace{-2mm}
& & & & & & & & & & &\\
\hline
&             &       &     &    &      &    &       &    && & \\
 &$I$ &5.27 &2.39 &2.30 &5.80 &3.80&3.72 &6.28 &3.86&4.87&2.38 \\
\vspace{-2mm}
         &$II$ &4.94 &2.50 &2.44 &5.56 &3.73&3.67 &6.00 &3.71&4.77&2.50 \\
\vspace{-2mm}
  $ZB$  &             &       &     &    &      &    &       &    && & \\
          &$III$   &6.05 &3.12 &3.00&5.78 &4.36&4.24 &6.24 &4.80&4.98&3.10  \\
     &$IV$  &5.07 &2.90 &2.82 &4.93 &3.83&3.75 &5.29 &4.14&4.36&2.90 \\
\hline
& & & & & & & & & & & \\
&&$E$_{\rm g}^{\Gamma-\Gamma}&$E$_{\rm g}^{\rm M-M}&$E$_{\rm g}^{\rm K-K}
&$E$_{\rm g}^{\rm \Gamma-M}&$E$_{\rm g}^{\rm \Gamma-K}&$E$_{\rm g}^{\rm A-A}
&$E$_{\rm g}^{\rm {M-\Sigma}}&$E$_{\rm g}^{\rm M-K}&$E$_{\rm g}^{\rm M-\Gamma}&$E$_{\rm g}^{\rm M-A}\\

\vspace{-2mm}
& & & & & & & & & & & \\
\hline
&             &       &     &    &      &    &       &    && & \\
&$I$       &5.11 &2.96 &3.53 &3.92 &4.17&6.42 &2.79&3.21&4.15&5.20 \\
\vspace{-2mm}
&$II$        &4.75 &3.05 &3.51 &3.85 &3.24&6.08 &2.88&~3.24&3.94&5.05 \\
\vspace{-2mm}
  $WZ$  &             &       &     &    &      &    &       &    && & \\

    &$III$        &5.22 &3.47 &3.76 &4.29 &4.36&6.16 &3.29&3.54&4.39&5.12 \\
&$IV$       &4.51 &3.23 &3.44 &3.81 &3.85&5.24 &3.08&3.26&3.93&4.49 \\
 \hline
\end{array}
$$
\end{table}

\section{Summary and conclusion}

A comparative study by FP-LAPW calculations based on DFT within
LDA, PBE-GGA, EV$_{\rm ex}$-PW$_{\rm co}$-GGA, and EV$_{\rm
ex}$-GGA-LDA$_{\rm co}$ schemes is introduced for the structural
and electronic properties of ScN in RS, CsCl, ZB, and WZ phases.
According to all approximations used in this work, the RS phase is
a stable ground state structure and makes a transition to CsCl
phase at high transition pressure. It can be concluded that
EV$_{\rm ex}$-PW$_{\rm co}$-GGA is the best one among the others
to provide accurate structural features and non-zero, positive
indirect band gap of RS-ScN comparable with the experimental
results when the structural and electronic calculations are aimed
to be calculated by the same exchange-correlation energy
approximation. Although LDA and PBE-GGA's have calculated good
structural features, they have yielded a small negative indirect
band gap for RS-ScN. On the other hand, EV$_{\rm
ex}$-GGA-LDA$_{\rm co}$ has roughly supplied the lattice constant
and bulk modulus but it is found to be very accurate for the
electronic features of RS-ScN. The indirect band gap of ScN in RS
phase is enlarged to the corresponding measured value by EV$_{\rm
ex}$-PW$_{\rm co}$-GGA+U$^{\rm SIC}$ calculations in which the
Coulomb self- and exchange-correlation interactions of the
localized $d$-orbitals of Sc have been corrected by the potential
parameter of U. The present EV$_{\rm ex}$-PW$_{\rm co}$-GGA
calculations have also provided good results for the structural
and electronic features of ZB, WZ and CsCl phases comparable with
the theoretical data reported in the literature. Therefore,
EV$_{\rm ex}$-PW$_{\rm co}$-GGA and EV$_{\rm ex}$-PW$_{\rm
co}$-GGA+U$^{\rm SIC}$ can be considered to be good
exchange-correlation energy approximations for further works of
ScN.

\ukrainianpart

\title
{Порівняльне дослідження структурних і електронних властивостей монокристалічного ScN}
\author{Р. Могаммад\refaddr{label1,label2},
        С. Катірчіоглу\refaddr{label2}}

\addresses{
\addr{label1} Палестинський технічний університет, коледж прикладних наук, ВестБенк, Палестина
\addr{label2} Близькосхідний технічний університет, фізичний факультет,
06530~Анкара, Туреччина}
\makeukrtitle

\begin{abstract}
\tolerance=3000%
Представлено порівняльне дослідження за допомогою  FP-LAPW розрахунків, що базуються на теорії функціоналу густини (DFT) в рамках схем LDA, PBE-GGA,
EV$_{\rm ex}$-PW$_{\rm co}$-GGA, і EV$_{\rm
ex}$-GGA-LDA$_{\rm co}$  для структурних і електронних властивостей ScN
в фазах  RS, ZB, WZ і CsCl. Відповідно до всіх наближень, виконаних в цій роботі, фаза RS є стійкою структурою в основному стані і здійснює перехід  у фазу
CsCl при високому тиску.  В той час  як схеми  PBE-GGA і EV$_{\rm
ex}$-PW$_{\rm co}$-GGA  забезпечують кращі структурні властивості такі як рівноважна постійна ґратки і об'ємні модулі, лише схеми
EV$_{\rm ex}$-PW$_{\rm co}$-GGA та EV$_{\rm ex}$-GGA-LDA$_{\rm
co}$'s дають ненульову, позитивну непряму енергетичну щілину для RS-ScN, порівняльну з експериментальними. Непряма зонна щілина  ScN в фазі RS
є збільшена до відповідного виміряного значення за допомогою  EV$_{\rm ex}$-PW$_{\rm co}$-GGA+U$^{\rm SIC}$ обчислень, в яких  кулонівські власні і
обмінні кореляційні  взаємодії
локалізованих  d-орбіталей Sc були поправлені за допомогою параметра потенціалу  U.  Розрахунки  EV$_{\rm ex}$-PW$_{\rm
co}$-GGA також приводять до добрих результатів для структурних і електронних характеристик  ScN у фазах ZB, WZ і CsCl, якщо порівнювати з теоретичними
даними, наявними в літературі. Вважається, що схеми EV$_{\rm ex}$-PW$_{\rm co}$-GGA та EV$_{\rm
ex}$-PW$_{\rm co}$-GGA+U$^{\rm SIC}$ є найкращими  серед інших у випадку, коли обчислюються структурні та електронні характеристики  ScN в рамках тих же
наближень для енергії обмінної кореляції.
\keywords ScN, FP-LAPW, DFT, структурні властивості, електронні властивості
\end{abstract}


\end{document}